\documentclass[letterpaper]{article} 
\usepackage{bm}
\usepackage{aaai2026}  
\usepackage{times}  
\usepackage{helvet}  
\usepackage{courier}  
\usepackage[hyphens]{url}  
\usepackage{graphicx} 
\usepackage{natbib}  
\usepackage{caption} 
\usepackage{algorithm}
\usepackage{algorithmic}
\usepackage{multirow}

\usepackage{color}
\usepackage{makecell} 
\definecolor{rred}{RGB}{245, 152, 153}
\definecolor{oorange}{RGB}{253, 205, 154}
\definecolor{carolinablue}{rgb}{0.6, 0.73, 0.89}
\usepackage{newfloat}
\usepackage{listings}
\usepackage{amssymb}
\usepackage{amsmath} 
\usepackage{bm}
\DeclareCaptionStyle{ruled}{labelfont=normalfont,labelsep=colon,strut=off} 
\floatstyle{ruled}
\newfloat{listing}{tb}{lst}{}
\floatname{listing}{Listing}
\title{RealisVSR: Detail-enhanced Diffusion for Real-World 4K Video Super-Resolution}
\author{
    Weisong Zhao\textsuperscript{\rm 1,\rm 2,6,\equalcontrib}, 
    Jingkai Zhou\textsuperscript{\rm 6,\equalcontrib}, 
    Xiangyu Zhu\textsuperscript{\rm 3,\rm 4}, \\
    Weihua Chen\textsuperscript{\rm 6, \thanks{Corresponding Author}}, 
    Xiao-Yu Zhang\textsuperscript{\rm 1,\rm 2,\footnotemark[2]}, 
    Zhen Lei\textsuperscript{\rm 3,\rm4,\rm 5,\footnotemark[2]},
    Fan Wang\textsuperscript{\rm 6}
}

\affiliations{
        \textsuperscript{\rm 1} Institute of Information Engineering, Chinese Academy of Sciences,\\
        \textsuperscript{\rm 2} School of Cyber Security, University of Chinese Academy of Sciences,\\
    \textsuperscript{\rm 3} State Key Laboratory of Multimodal Artificial Intelligence Systems, Institute of Automation, Chinese Academy of Science,\\
    \textsuperscript{\rm 4} School of Artificial Intelligence, University of Chinese Academy of Sciences,\\
    \textsuperscript{\rm 5}CAIR, Hong Kong Institute of Science \& Innovation, Chinese Academy of Sciences,
    \textsuperscript{\rm 6} DAMO Academy\\
\vspace{3mm}
\setcounter{figure}{0} 
\includegraphics[width=\linewidth]{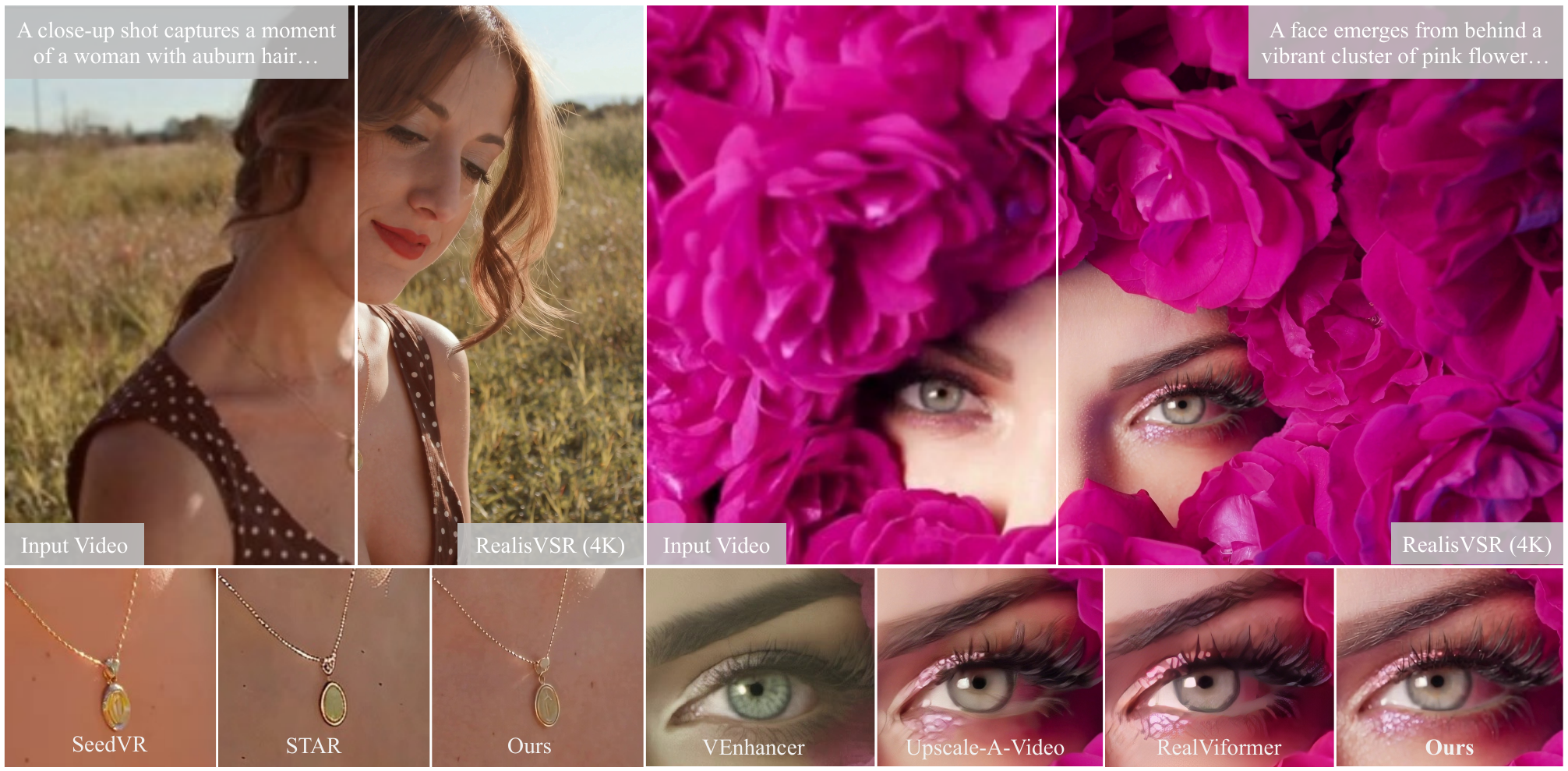}
\vspace{-4mm}
\captionof{figure}{Visual comparisons of RealisVSR with the leading VSR methods in 4K-target video super-resolution scenarios. Compared to the state-of-the-art VSR models, \emph{e.g.,} Upscale-A-Video \cite{uav}, RealViformer \cite{realviformer}, STAR \cite{star}. We achieve more sophisticated detail enhancement and better consistency. \textbf{(Zoom-in for best view)} \label{fig:teaser}
}
}

\usepackage{bibentry}

\begin{document}

\maketitle

\begin{abstract}
Video Super-Resolution (VSR) has achieved significant progress through diffusion models, effectively addressing the over-smoothing issues inherent in GAN-based methods. Despite recent advances, three critical challenges persist in VSR community: 1) Inconsistent modeling of temporal dynamics in foundational models; 2) limited high-frequency detail recovery under complex real-world degradations; and 3) insufficient evaluation of detail enhancement and 4K super-resolution, as current methods primarily rely on 720P datasets with inadequate details.
To address these challenges, we propose \textbf{RealisVSR}, a high-frequency detail-enhanced video diffusion model with three core innovations: 
1) Consistency Preserved ControlNet (CPC) architecture integrated with the Wan2.1 video diffusion to model the smooth and complex motions and suppress artifacts; 
2) High-Frequency Rectified Diffusion Loss (HR-Loss) combining wavelet decomposition and HOG feature constraints for texture restoration; 
3) \textbf{RealisVideo-4K}, the first public 4K VSR benchmark containing 1,000 high-definition video-text pairs.
Leveraging the advanced spatio-temporal guidance of Wan2.1, our method requires only 5–25\% of the training data volume compared to existing approaches.
Extensive experiments on VSR benchmarks (REDS, SPMCS, UDM10, YouTube-HQ, VideoLQ, RealisVideo-720P) demonstrate our superiority, particularly in ultra-high-resolution scenarios. (https://zws98.github.io/RealisVSR-project/)
\end{abstract}

\section{Introduction}
Real-world video super-resolution (VSR) requires reconstructing high-resolution sequences with spatial precision and frame-to-frame coherence under unpredictable degradation conditions. 
Conventional VSR frameworks primarily address synthetic degradations (e.g., idealized downsampling or sensor-specific noise), practical applications encounter heterogeneous degradation patterns combining compression artifacts, motion blur, and sensor imperfections. 
These complex distortions disrupt temporal modeling in deep architectures, challenging simultaneous detail recovery and temporal consistency across sequences.

While adversarial learning frameworks \cite{realesrgan,realbasicvsr} remain prevalent due to their texture-enhancing capabilities, their reliance on optical flow for temporal alignment often falls short in real-world scenarios. 
Such methods typically produce artifacts despite generating motion-smoothed outputs. 
Emerging approaches leverage diffusion priors \cite{sd} and show promise for photorealistic restoration, with extensions like \cite{uav, seedvr} incorporating spatiotemporal layers to improve coherence. 
However, their fundamental models pretrained in image distribution create a mismatch for video tasks: these models require extensive video data to learn motion dynamics from scratch, yet still exhibit temporal inconsistencies. 
STAR \cite{star} adopt CogVideoX-5B \cite{cogvideox} to exemplify this limitation, which demands about 200K training samples due to inadequate temporal modeling capacity in the fundamental model. 
To overcome this, we propose a VSR framework built on Wan2.1 \cite{wan2.1}, a video diffusion model demonstrated to achieve superior temporal consistency. 
Specifically, we introduce Consistency Preserved ControlNet (CPC) architecture that omits noisy latent injections in condition branch to suppress the occurrence of artifacts. 
Notably, our method achieves state-of-the-art performance while requiring only 25\% of STAR's training data.

Moreover, reconstructing high-frequency details (e.g., sharp edges, fine textures) remains critical for perceptually compelling VSR. 
Current diffusion-based methods predominantly employ rectified flow losses \cite{rectifiedflow} to predict velocity fields. 
Although their effectiveness for pixel-level reconstruction in image tasks, these losses lack explicit frequency-domain discrimination, leading to suboptimal high-frequency recovery as networks prioritize low-frequency structural fidelity. 
To address this, we propose a High-Frequency Rectified Loss (HR-Loss) to enhance high-frequency components. 
Our analysis of wavelet-based and HOG-based features demonstrates that rectifying the high-frequency components of flow significantly improves fine details and textures in generated videos.

Despite progress, public VSR benchmarks (e.g., REDS \cite{reds}) and state-of-the-art methods focus on upscaling to 720P resolutions, inherently limiting the evaluation of ultra-fine detail recovery (e.g., hair strands, fabric textures). 
The absence of open-source ultra-high-resolution (4K) datasets forces reliance on no-reference metrics (e.g., NIQE \cite{niqe}, DOVER \cite{dover}) for ultra-high-resolution scenarios. 
Such evaluations overlook critical texture fidelity and create a validation gap for high-frequency restoration techniques, leading to potential biases. 
To bridge this gap, we advocate for an open-access 4K VSR dataset with 1,000 professionally captured ground-truth videos in diverse scenes, enabling rigorous assessment and algorithmic innovation. 

To sum up, our contributions are summarized as follows:
\begin{itemize}
    \item We propose a novel VSR framework leveraging Wan2.1 as the video diffusion prior, featuring a CPC architecture that eliminates injected noise in conditional inputs to suppress artifacts. 
    To the best of our knowledge, this is the first work to deploy a VSR model on Wan2.1.
    \item We introduce an HR-Loss that explicitly enhances high-frequency details through wavelet decomposition and HOG-based modeling, mitigating the low-frequency bias in rectified flow losses.
    \item We establish RealisVideo-4K, the first detail-rich and open-access 4K dataset including 1,000 4K video-text pairs, facilitating realistic VSR evaluation.
    \item Extensive experiments validate the superiority of RealisVSR across benchmarks.
\end{itemize}

\section{Related Work}
\subsection{Open-source Text-to-video Model}
Recent advancements in text-to-video (T2V) generation have been driven by both closed-source commercial models and open-source initiatives. Closed-source models like OpenAI’s Sora \cite{sora}, Runway’s Gen-3 \cite{gen3}, Kling \cite{kling}, and Google’s Veo 2 \cite{veo2} demonstrate exceptional quality but remain inaccessible to the broader community. Open-source efforts, such as HunyuanVideo \cite{hunyuanvideo}, Mochi \cite{mochi}, and CogVideoX \cite{cogvideox}, have narrowed this gap by releasing foundational architectures and training frameworks. However, these models often face three key limitations: suboptimal performance compared to state-of-the-art closed systems, narrow task scope limited to basic T2V, and inefficiency for resource-constrained environments. For instance, many open models require high-end GPUs due to memory-intensive designs, hindering their adoption in consumer-grade applications. Additionally, challenges in generating coherent motion and long-term temporal consistency persist across existing frameworks. 
Wan2.1 \cite{wan2.1} emerges as a breakthrough in open-source T2V generation, addressing these limitations through architectural innovations and scalable training strategies. Built on a diffusion transformer (DiT) framework enhanced with spatio-temporal attention and a novel variational autoencoder (Wan-VAE), Wan2.1 achieves superior performance across benchmarks like Wan-Bench and human preference evaluations. Its dual-parameter variants Wan1.3B and Wan14B set new standards in motion complexity and physical plausibility. Key advantages include its ability to preserve fine-grained details and dynamic scene understanding via temporal coherence optimization. These features make Wan2.1 uniquely suited for enhancing low-resolution videos while maintaining the temporal consistency of generated videos. 

\subsection{Video Super-resolution}
Video super-resolution (VSR) aims to reconstruct high-resolution (HR) video sequences from degraded low-resolution (LR) inputs, with a focus on handling real-world complexities such as motion blur and sensor noise. GAN-based methods have dominated recent advancements, leveraging adversarial training to recover fine textures and enhance perceptual quality. Notable works like RealBasicVSR \cite{realbasicvsr} and RealViformer \cite{realviformer} integrate pre-cleaning modules and channel attention mechanisms to address unknown degradations. However, these approaches often struggle with over-smoothing high-frequency details and hallucinating artifacts under stochastic distortions \cite{uav, star}.

Recent works have explored text-to-video (T2V) diffusion priors for VSR tasks, leveraging large-scale pre-trained models to capture rich spatio-temporal prior knowledge. DAM-VSR \cite{damvsr} builts on a two-stage pipeline, which requires an initial image super-resolution module to produce a high-quality reference image before video super-resolution framework. In contrast, the mainstream methods prefer the end-to-end pipeline. Upscale-A-Video \cite{uav}, MGLD-VSR \cite{mgldvsr} and SeedVR \cite{seedvr} extend T2V architectures by incorporating temporal layers for motion-aware super-resolution. However, it inherits limitations from text-to-image diffusion paradigms, leading to temporal incoherence due to static noise schedules and inadequate motion modeling. More recent approaches like VEnhancer \cite{venhancer} applies T2V priors to AI-generated videos but demonstrate limited generalization under real-world degradations, such as sensor noise and non-uniform blur. STAR \cite{star} finetunes CogVideoX \cite{cogvideox} on a 720P video dataset. However, it demands 200K training samples due to inherent limitations in the base model's motion modeling, and the videos used for supervision lack high-frequency details, leading to inferior super-resolution performance.

\section{Method}
\subsection{Overall architecture of RealisVSR}
We build our video super-resolution method, RealisVSR, on the leading Wan2.1 model. We adopt ControlNet \cite{controlnet, cogvideox} architecture to inject the low-quality video condition information in the denoising process of RealisVSR. There are two basic modules in RealisVSR: 1) fundamental model including 3D Causal VAE, DiT blocks and Text Encoder; 2) Consistency Preserved ControlNet. 
\subsubsection{Fundamental Model}
First, we take Wan-VAE as a causal spatio-temporal encoder-decoder architecture for video compression. Specifically, given a video input $\mathcal{V} \in \mathbb{R}^{(1+T) \times H \times W \times 3}$, the Wan-VAE encoder maps it to compressed latent dimensions $[1 + T/4, H/8, W/8]$ with channel expansion from $C=3$ to $C'=16$. Notably, the initial frame $\mathcal{V}_0$ undergoes exclusive spatial compression ($H \times W \rightarrow H/8 \times W/8$) to preserve spatial fidelity for image-domain conditioning. The DiT block \cite{dit} comprises three key components: a patchifying module using 3D convolution to map inputs to spatio-temporal tokens $(B, L, D)$, transformer blocks with cross-attention for text-conditioning and shared MLPs for time embedding modulation, and an unpatchifying module for reconstruction. 
For text encoder, Wan 2.1 employs umT5 \citep{umt5} to leverage its strong multilingual capabilities for simultaneous Chinese/English understanding and visual text processing. 
\begin{figure}[t]
\centering 
\includegraphics[width=\linewidth]{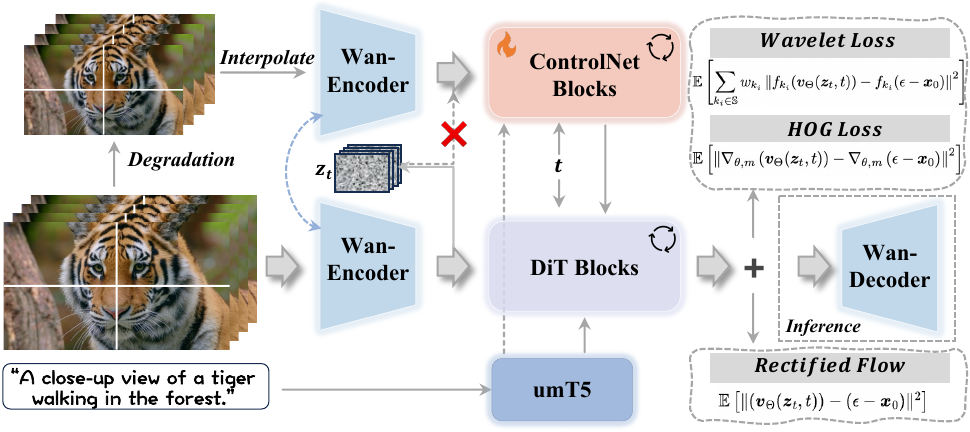}
\caption{The framework of the proposed RealisVSR.}
\label{fig1}
\end{figure}
\subsubsection{Consistency Preserved ControlNet}
Here we describe the proposed Consistency Preserved ControlNet, focusing on the removal of noisy latent input $z_t$ in the ControlNet block while maintaining convergence properties. The vanilla ControlNet \cite{controlnet, cogvideox} integrates both the noisy latent $z_t\in \mathbb{R}^{B\times C \times T \times H \times W}$ and conditional signals through adaptive normalization layers, \emph{i.e.,} $x_{\text{comb}} = \mathcal{T}_{\theta_{\text{patch}}}(z_t) + \mathcal{T}_{\theta_{\text{cond}}}(c_t)$
. Our key innovation lies in discarding $z_t$ at the input stage, \emph{i.e.,} $x_{\text{CPC}} =\mathcal{T}_{\theta_{\text{cond}}}(c_t)$
, which surprisingly improves visual quality by suppressing artifacts while preserving training stability. Then, the CPC features are fused into the main network through a depth-adaptive residual connection mechanism. Given $L_{\text{main}}$ main network blocks and $L_{\text{CPC}}$ CPC blocks, we compute the depth ratio $r = \lceil L_{\text{main}} / L_{\text{CPC}} \rceil$ to align feature resolutions across architectures. At each main network block $i$, the corresponding CPC feature $\mathbf{F}_{\lfloor i/r \rfloor}^{\text{CPC}}$ is added to the intermediate output with learnable scaling:
\begin{equation}
\mathbf{X}_i^{\text{main}} \leftarrow \mathbf{X}_i^{\text{main}} + \gamma \cdot \mathbf{F}_{\lfloor i/r \rfloor}^{\text{CPC}},
\label{eq:1}
\end{equation}
where $\gamma$ denotes the scaling factor and $\lfloor \cdot \rfloor$ represents floor division. This enables multi-scale conditioning while maintaining compatibility with varying network depths.

\subsection{High-frequency Rectified Loss}
We present the High-Frequency Rectified Loss (HR-Loss) to explicitly prioritize high-frequency component restoration while preserving low-frequency structural coherence. The proposed HR-Loss achieves enhancement of fine-grained details and textural fidelity in video super-resolution tasks. The diffusion process can be formulated as follows:
\begin{equation}
z_t = \alpha_t \cdot x_0+\sigma_t \cdot \epsilon,
\label{eq:2}
\end{equation}
where $x_0, \epsilon$ are real video distribution and standard normal distribution, respectively. $\alpha_t$ and $\sigma_t$ are factors in the diffusion formulation. SD3 \cite{sd3} and Flux \cite{flux} introduce rectified flow loss to predict velocity $\bm{v}$ parameterized by $\Theta$:
\begin{equation}
\mathcal{L}_{\text{REC}} = \mathbb{E} \left[ \left\|  \bm{v}_\Theta(\bm{z}_t, t) - \epsilon - \bm{x}_0  \right\|^2 \right],
\label{eq:3}
\end{equation}
To further strengthen detail recovery, we extend the descriptors commonly used in pixel space to latent space,  as many works have found that some pixel operations are also applicable in latent space, e.g., Demofusion \cite{demofusion} applies pixel space interpolation to latent space, Video-P2P \cite{videop2p} and Image-P2P \cite{imagep2p} apply front/back background stitching to latent space.
We incorporate two high-frequency modeling strategies:

\textbf{1) Wavelet-based Frequency Decomposition:}  
Building on the diffusion process $\bm{z}_t = \alpha_t \bm{x}_0 + \sigma_t \epsilon$ (Eq.~\ref{eq:2}), we decompose latent features into multi-scale sub-bands using Haar wavelet transforms. For a velocity prediction model $\bm{v}_\Theta(\bm{z}_t, t)$, the wavelet loss is formulated as:
\begin{equation}
\mathcal{L}_{\text{WLF}} = \mathbb{E} \left[ \sum_{k_i \in \mathbb{S}} w_{k_i} \left\| f_{k_i}(\bm{v}_\Theta(\bm{z}_t, t)) - f_{k_i}(\epsilon - \bm{x}_0) \right\|^2 \right],
\label{eq:4}
\end{equation}
where $\mathbb{S} = \{\text{LL, LH, HL, HH}\}$ denotes sub-band components, $f_{k_i}(\cdot)$ applies wavelet kernels (low-pass $L = \frac{1}{\sqrt{2}}[1, 1]$, high-pass $H = \frac{1}{\sqrt{2}}[-1, 1]$), and $w_{f_i}$ weights prioritize high-frequency bands (LH, HL, HH). We set the weight set $\mathbf{w}=\{w_{LL},w_{LH},w_{HL},w_{HH}\}=\{1.0,2.0,2.0,2.0\}$.

\textbf{2) HOG-based Texture Constraint:}  
We introduce a Histogram of Oriented Gradients (HOG) feature loss. Specifically, we collect HOG feature \cite{hog} for the output latent in each latent channel. For gradient orientations $\theta$ discretized into 9 bins spanning $[0^\circ, 180^\circ]$ (each bin covers $20^\circ$) and magnitudes $m$ normalized via L2-Hys regularization, the HOG loss penalizes discrepancies in directional texture patterns:
\begin{equation}
\mathcal{L}_{\text{HOG}} = \mathbb{E} \left[ \left\| \nabla_{\theta, m} \left( \bm{v}_\Theta(\bm{z}_t, t) \right) - \nabla_{\theta, m} \left( \epsilon - \bm{x}_0 \right) \right\|^2 \right],
\label{eq:5}
\end{equation}
where $\nabla_{\theta, m}(\cdot)$ computes gradient histograms over spatial neighborhoods. The final loss combines both frequency-specific constraints and rectified flow loss:
\begin{equation}
\mathcal{L}_{\text{HR}} = \mathcal{L}_{\text{WLF}} + \mathcal{L}_{\text{HOG}} + \mathcal{L}_{\text{REC}},
\label{eq:6}
\end{equation}
where $\mathcal{L}_{\text{REC}} = \|\bm{v}_\Theta(\bm{z}_t, t) - (\epsilon - \bm{x}_0)\|^2$ preserves structural fidelity. This enables RealisVSR to recover ultra-fine textures while maintaining global coherence.
\subsection{RealisVideo-4K}
Current popular VSR benchmarks, \emph{e.g.,} REDS \cite{reds}, SPMCS \cite{spmcs}, UDM10 \cite{udm10}, Youtube-HQ \cite{realbasicvsr} and Vid4 \cite{vid4}, are collected at 720P resolution and evaluated via full-reference metrics, \emph{e.g.,} PSNR, SSIM, LPIPS \cite{lpips} and DIST \cite{dist}. However, the ground truth videos at this resolution lack rich details, which impact the evaluation of methods on detail enhancement. Some no-reference metrics, \emph{e.g.,} DOVER \cite{dover} score and CLIP-IQA \cite{clipiqa} score are presented to measure the general quality of videos. However, we observe that the inconsistent output video with additional artifacts may inevitably increase these no-reference metrics. Therefore, to explore the capacity of models for detail enhancement, a high-resolution dataset with rich details is needed. We collect and filter about 1,000 4K videos containing rich details and randomly select 140 detail-rich videos to build the testing set. The filtering strategy is two-stage. In the first stage, we use QWen2.5-VL \cite{qwen2.5vl} to annotate the videos with text, and utilize some keywords, such as ``close-up", ``detailed", etc., to select the videos with rich details. Then, we manually filter out the low-quality videos that are wrongly selected due to wrong text annotation. 

In addition, we also construct a 720P version by resizing the videos of RealisVideo-4K to 720P resolution to align with the existing benchmarks, named RealisVideo-720P. Unlike existing datasets that use inherently limited 720P ground-truth videos, the proposed RealisVideo-720P is derived from authentic 4K source material downsampled to 720P via interpolation, preserving subpixel-level details lost in native 720P benchmarks. This design enables rigorous evaluation of super-resolution models' ability to recover high-frequency textures and structural fidelity, offering a more realistic and challenging assessment of detail restoration capabilities. To our knowledge, RealisVideo is the first 4K detail-rich video dataset for VSR tasks.

\begin{table*}[h]
    \begin{center}
    \renewcommand{\arraystretch}{1.2}
    \resizebox{\linewidth}{!}{
    \begin{tabular}{c|c|c|c|c|c|c|c|c|c}
    \hline
    \multirow{2}{*}{\textbf{Datasets}} &  \multirow{2}{*}{\textbf{Detail Level}} &  \multirow{2}{*}{\textbf{Metric}} & \textbf{Upscale-A-Video} & \textbf{VEnhancer} & \textbf{RealViformer} & \textbf{MGLD-VSR} & \textbf{SeedVR} & \textbf{STAR}   & \multirow{2}{*}{\textbf{RealisVSR (Ours)}} \\
    \cline{4-9}
    & & & CVPR'24 & Arxiv'24 & ECCV'24 & ECCV'24 & CVPR'25 & ICCV'25 &\\
    \hline
    \multirow{4}{*}{VideoLQ} & \multirow{4}{*}{Low} & DOVER $\textcolor{red}{\uparrow}$	& 36.37 & 50.41 & 44.23 & 48.73 & 46.59	& \underline{51.28} & \textbf{51.96}	\\
        & & MUSIQ $\textcolor{red}{\uparrow}$	& 35.65	& \textbf{64.40}	& \underline{63.47} &59.45& 55.19 & 58.75 &	56.40		 \\
        & & NIQE $\textcolor{green}{\downarrow}$ 	& 5.89 &	4.36 &	5.32 & \textbf{3.69} & \underline{4.33} & 4.47 &	5.18	\\
    \hline
    \multirow{5}{*}{SPMCS} & \multirow{20}{*}{Medium} & PSNR $\textcolor{red}{\uparrow}$ & 24.70 & 16.96 & \underline{25.60} & 22.92 & 24.31 & 23.35 & \textbf{27.36} \\
        & & SSIM $\textcolor{red}{\uparrow}$ & 0.6689 & 0.4334 & \underline{0.7601} & 0.6327 &	0.7257 & 0.6747 & \textbf{0.8169} \\
        & & LPIPS $\textcolor{green}{\downarrow}$& 0.3720 & 0.4307 &	0.3121 & 0.2861 & \underline{0.2165} & 0.2670 & \textbf{0.1388}  \\
        & & DISTS $\textcolor{green}{\downarrow}$ & 0.2094 & 0.1891 & 0.1906 & 0.1419 & \underline{0.0970} & 0.1209 & \textbf{0.0780} \\
       &  & DOVER $\textcolor{red}{\uparrow}$  & 36.06 & 48.53 & 46.65 &44.77&	51.43 & \textbf{52.63} & \underline{51.68}  \\
    \cline{1-1}\cline{3-10}
    \multirow{5}{*}{UDM10} & & PSNR $\textcolor{red}{\uparrow}$ & 25.69	& 18.61 & \underline{28.74}	& 25.97 &	25.68 & 25.73	& \textbf{29.23}	\\
       &  & SSIM $\textcolor{red}{\uparrow}$ 	& 0.7556 &	0.5610 & \underline{0.8505} & 0.7572 &	0.7753 & 0.7907 &	\textbf{0.8708}  \\
       &  & LPIPS $\textcolor{green}{\downarrow}$  &	0.3240 & 0.3907	 &	0.2036 &0.2249&	\underline{0.1915} & 0.2061 &	\textbf{0.1221}  \\
       &  & DISTS $\textcolor{green}{\downarrow}$  &	0.1835 & 0.1599 &	0.1454 & 0.1160 &	\underline{0.0845} & 0.0978 &	\textbf{0.0683}  \\
       &  & DOVER $\textcolor{red}{\uparrow}$  &	43.71 &	58.00 &	53.15 & 54.65 &	57.91 & \textbf{62.54} &	\underline{58.93}  \\
    \cline{1-1}\cline{3-10}
    \multirow{5}{*}{REDS30}& & PSNR $\textcolor{red}{\uparrow}$ 	& {22.95}	& 18.41 &	\underline{25.43}	& 23.38 &	24.61 & 23.52	& \textbf{26.56}	 \\
        & & SSIM $\textcolor{red}{\uparrow}$ 	& 0.5836 &	0.4559 & 0.7134 & 0.6223 &	\underline{0.7216} & 0.6393 &	\textbf{0.7593} \\
        & & LPIPS $\textcolor{green}{\downarrow}$ &	0.4528 &	0.4536 &	0.2628 & 0.2348 &	\textbf{0.2234} & 0.2710 &	\underline{0.2500} \\
        & & DISTS $\textcolor{green}{\downarrow}$ &	0.2339 &	0.1597 &	0.1455 &0.1010&	\textbf{0.0882} & 0.0972 &	\underline{0.0991} \\
       &  & DOVER $\textcolor{red}{\uparrow}$  &	34.37 &	39.71 &	41.39 & 42.83 &	39.58 & \textbf{48.64}&	\underline{44.86}  \\
    \cline{1-1}\cline{3-10}
    \multirow{5}{*}{YouTube-HQ}& & PSNR $\textcolor{red}{\uparrow}$	& 23.44	& 17.38 &	\underline{24.45}	& 23.42 &	23.68 & 23.35	& 
    \textbf{25.84}	\\
        & & SSIM $\textcolor{red}{\uparrow}$ 	& 0.6138 &	0.4345 &	\underline{0.7264} & 0.6237 &	0.6721 & 0.6747 &	\textbf{0.7601} \\
        & & LPIPS $\textcolor{green}{\downarrow}$ &	0.3861 & 0.4393 &	0.2667 & 0.2835 &	\underline{0.2131} & 0.2660 &	\textbf{0.1935}  \\
        & & DISTS $\textcolor{green}{\downarrow}$  &	0.2156 &	0.1716 &	0.1907 & 0.1463 &	\textbf{0.0809} & 0.1209 &	\underline{0.0853} \\
       
       &  & DOVER $\textcolor{red}{\uparrow}$  &	52.98 &	65.50 &64.63 & 61.78 & \underline{70.11} & \textbf{71.38} &	64.67  \\
    \hline
    \multirow{5}{*}{\makecell{RealisVideo\\-720P}} & \multirow{5}{*}{High} & PSNR $\textcolor{red}{\uparrow}$	& 26.81	& 20.04 &	\underline{27.77}	& 26.33 &	26.97 & 26.31	& \textbf{28.23}	\\
        & & SSIM $\textcolor{red}{\uparrow}$ 	& 0.7710 &	0.6255 &	\underline{0.8342} & 0.7556 &	0.8002 & 0.7860 &	\textbf{0.8460} \\
        & & LPIPS $\textcolor{green}{\downarrow}$ &	0.2712 & 0.3727 &	\underline{0.1501} & 0.2470 &	0.1850 & 0.1856 &	\textbf{0.1414}  \\
        & & DISTS $\textcolor{green}{\downarrow}$  & 0.1798 &	0.1650 &	0.1179 & 0.1418 &	\underline{0.0940} & 0.1007 &	\textbf{0.0844} \\
       &  & DOVER $\textcolor{red}{\uparrow}$  &	58.69 &	68.60 &	67.96 & 63.57 &	70.36 & \underline{74.96} &	\textbf{77.34}  \\
    \hline
    \multirow{5}{*}{\makecell{RealisVideo\\-4K}} & \multirow{5}{*}{\makecell{Very\\High}} & PSNR $\textcolor{red}{\uparrow}$	& 29.13	& 20.59 &	\underline{30.01}	& 26.26 &	24.73 & 26.52	& \textbf{30.90}	\\
        & & SSIM $\textcolor{red}{\uparrow}$ 	& \underline{0.8592} &	0.7391 &	0.8428 & 0.7898 &	0.7847 & 0.8052 &	\textbf{0.8746} \\
        & & LPIPS $\textcolor{green}{\downarrow}$ &	0.2310 & 0.3610 &	\underline{0.1355} & 0.2329 &	0.2270 & 0.2331 &	\textbf{0.0939}  \\
        & & DISTS $\textcolor{green}{\downarrow}$  & 0.1690 &	0.1932 &	\underline{0.1597} & 0.1988 &	0.1352 & 0.1691 &	\textbf{0.1310} \\
       &  & DOVER $\textcolor{red}{\uparrow}$  &	60.66 &	67.98 &	68.09 & 63.85 &	70.61 & \textbf{72.72} &	\underline{70.06}  \\
    \hline
    \end{tabular}
    }
    \end{center}
    \vspace{-2mm}
        \caption{
        Quantitative comparisons on VSR benchmarks from diverse scenarios, \emph{i.e.,} synthetic (SPMCS, UDM10, REDS, YouTube-HQ), and real (VideoLQ) datasets. The best and second performances are marked in bold and underlined, respectively. The RealisVideo-720P is resized from RealisVideo-4K, including the richest details among 720P datasets. 
        }
         \vspace{-1mm}
    \label{tab:comparison}
\end{table*}
\section{Experiments}
\subsection{Dataset and Implementation Details}
\subsubsection{Training Data}
We train RealisVSR on a self-collected 4K dataset, containing merely 50K text-video pairs. During training, each video is randomly cropped to a 480P-scale resolution for training and inference efficiency. For this, we predefine a 480P resolution bucket including different aspect ratios. For each video from within that bucket, the resolution is sampled in order for video cropping. To construct the LR-HR video pairs, we follow the mainstream methods to adopt a two-order video degradation pipeline \cite{realbasicvsr}. 
\subsubsection{Testing Data}
We test our method on both synthetic and real datasets at 720P and 4K. 1) \textbf{720P:} For synthetic dataset, we follow the existing methods \cite{uav,star,seedvr,realbasicvsr,realviformer} to construct LR-HR pair with the same degradation pipeline in training for REDS \cite{reds}, SPMCS \cite{spmcs}, UDM10 \cite{udm10}, and Youtube-HQ \cite{realbasicvsr}. These datasets do not include rich details. Therefore, we randomly sample 140 videos from the proposed RealisVideo-4K and resize them to 720P and construct LR-HR pairs with the same degradation pipeline, and this paired dataset is named as RealisVideo-720P. For real dataset, we utilize Video-HQ \cite{uav} as the prior work adopted. 
2) \textbf{4K:} We adopt the proposed RealisVideo-4K, which contains 1,000 4K videos with rich details, from which 140 are randomly selected as the testing set. We construct LR-HR pairs with two-order degradation pipeline.
\subsubsection{Implementaion Details}
We adopt Wan-1.3B as our fundamental model and train Consistency Preserved ControlNet on it. The scale of condition is set to 1.0. We follow Wan2.1 to adopt FSDP \cite{fsdp} and set the sequence degree to 2. The model is trained on 8 NVIDIA H20-100G GPUs with 12K interations and a batch size of 8 via gradient accumulation strategy. The videos for training are cropped to a random size at 480P resolution and 49 frames with a predefined resolution bucket. We adopt AdamW \cite{adamw} as the optimizer. The weight decay is 1e-4 and the learning rate is 1e-5.
\paragraph{Evaluation Metrics}
For the synthetic datasets, we employ full-reference metrics such as PSNR, SSIM, LPIPS \cite{lpips} and DIST \cite{dist}. Since the real dataset Video-LQ does not include ground truth, we adopt no-reference metrics, \emph{e.g.,} DOVER \cite{dover}, NIQE \cite{niqe} and CLIP-IQA \cite{clipiqa}. However, we observe that a higher degree of artifacts can inflate the scores of no-reference metrics, thus compromising their reliability. Therefore, we focus more on full-reference metrics.
\begin{figure*}[!t]
\centering 
\includegraphics[width=\linewidth]{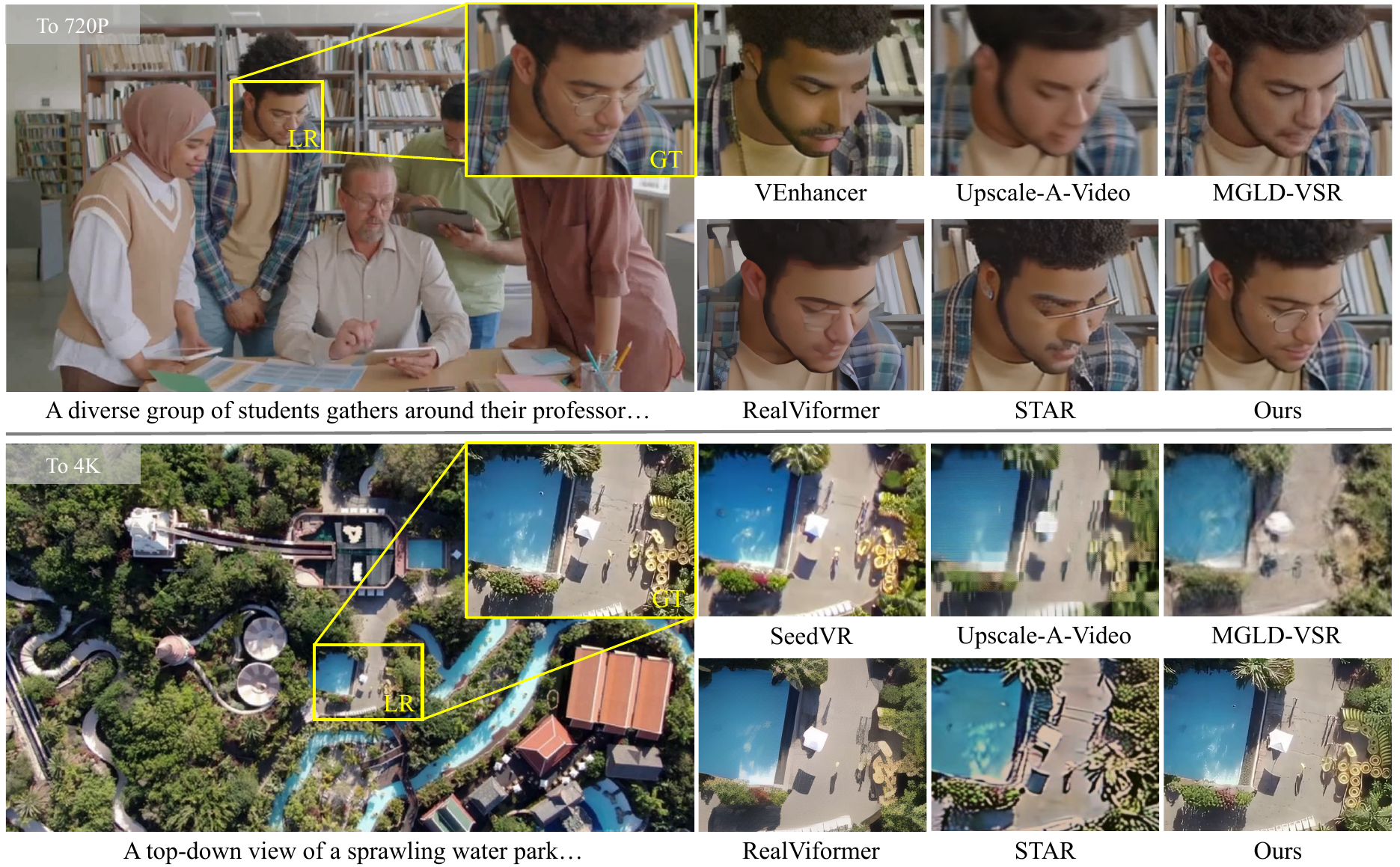}
\vspace{-5mm}
\caption{Qualitative assessments on a detail-rich dataset at 720P (first row) and 4K (second row) demonstrate that our RealisVSR generates highly realistic details. Compared to existing methods, it exhibits superior restoration capabilities, effectively eliminating degradations while preserving intricate textures—such as those of glasses and poolside clutter.}
\vspace{-2mm}
\label{fig:compare}
\end{figure*}
\subsection{Comparison at 720P}
\subsubsection{Quantitative Results}
As shown in Tab. \ref{tab:comparison}, we assess five metrics for each synthetic benchmarks, \emph{e.g.,} REDS, SPMCS, UDM10, Youtube-HQ, RealisVideo-720P, and compare with the leading methods, \emph{e.g.,} Upscale-A-Video \cite{uav}, VEnhancer \cite{venhancer}, RealViformer \cite{realviformer}, MGLD-VSR \cite{mgldvsr}, SeedVR \cite{seedvr} and STAR \cite{star}. Our RealisVSR achieves the best performance in most of these metrics. It is worth noting that MGLD-VSR and RealViFormer are trained on REDS, which shows strong performance on the corresponding test set. Even so, our approach outperforms them on this benmark. SeedVR and STAR achieve outstanding perceptual quality, but they cannot preserve the consistent output with origin input. By contrast, our method not only recovers rich details but maintain the consistent output. These outcomes underscore the capacity of our method to produce visually realistic details with strong fidelity. Additionally, when evaluated on a real-world dataset, \emph{e.g.,} Video-LQ, we utilize three no-reference metrics, \emph{e.g.,} DOVER, MUSIQ and NIQE. Our method achieves the best DOVER score and competitive MUSIQ and NIQE. Note that we observe that a greater degree of artifacts inevitably increases these no-reference metrics. Moreover, we evaluate the inference time and dataset size of different VSR methods in the Supplementary Materials. The results showcase the superior efficiency of our method.
\subsubsection{Qualitative Results}
Fig. \ref{fig:compare} shows results on detail-rich benchmarks with ground-truth,\emph{i.e.,} RealisVideo-720P. Our method outperforms existing VSR approaches by a large margin in both degradation removal and texture generation. Specifically, our method effectively recovers detailed structures, such as the glasses of young student. In addition, our method faithfully restores fine details where other approaches produce blurred and inconsistent details.

\begin{table}[!t]
    \begin{center}
    \renewcommand{\arraystretch}{1.2}
    \resizebox{\linewidth}{!}{
    \begin{tabular}{c|c|c|c|c|c}
    \hline
    \multirow{2}{*}{\textbf{Dataset}} & \multicolumn{5}{c}{\textbf{Warping Error($E_{warp}$$\textcolor{green}{\downarrow}$	)}} \\
     \cline{2-6}
      & \textbf{VEnhancer} & \textbf{RealViformer} & \textbf{MGLD} & \textbf{STAR}   & \textbf{Ours} \\
    \hline
     SPMCS 	& 1.40 & 1.32 & 2.57 &	2.00 & \textbf{0.81}	\\
     UDM10  & 2.47 & 2.13 & 3.29 & 2.90 & \textbf{1.72}	\\
     REDS & 1.35 & 2.45 & 2.78 &	2.51 & \textbf{0.83}	\\
     Youtube-HQ & 1.98 & 2.37 & 2.97 & 3.20 & \textbf{1.59}	\\

    \hline
    \end{tabular}
    }
    \end{center}
    \vspace{-2mm}
        \caption{
        Temporal consistency comparison with SOTA methods on warping error $E_{warp}$ \cite{warp}.
        }
         \vspace{-1mm}
    \label{tab:consistency}
\end{table}
\begin{figure*}[!t]
\centering 
\includegraphics[width=\linewidth]{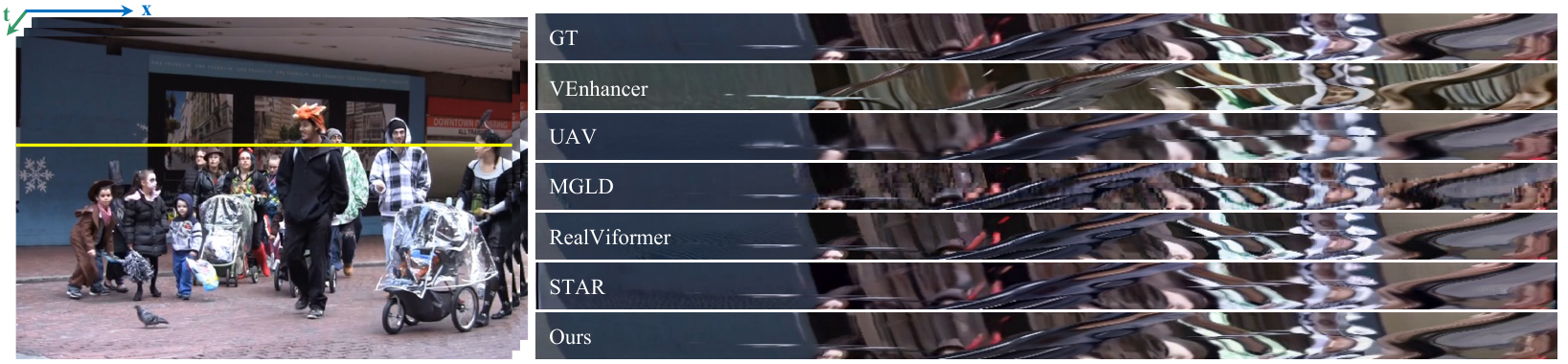}
\vspace{-3mm}
\caption{Qualitative comparison of temporal consistency with the existing video super-resolution methods. We assess a row and track changes over time on Vid4 \cite{vid4}. \textbf{(Zoom-in for best view)}}
\label{fig:flow}
\end{figure*}
\subsubsection{Temporal Consistency}
We conduct comprehensive evaluations of temporal consistency across four benchmark datasets—SPMCS, UDM10, REDS, and YouTube-HQ—using the warping error metric $E_{warp}$ \cite{warp} to quantify frame-to-frame motion coherence. The results in Tab. \ref{tab:consistency} demonstrate that our approach achieves state-of-the-art performance, with significant improvements over existing methods. This is assisted by the powerful modeling capacity of Wan2.1. On the SPMCS dataset, our method achieves a remarkably low warping error ($E_{warp}$) of 0.81, significantly surpassing the best diffusion-based competitor STAR ($E_{warp}$=2.00) by 60\% and the leading GAN-based approach RealViformer ($E_{warp}$=1.32) by 37\%, demonstrating a substantial leap in temporal fidelity. Similarly, on the challenging UDM10 dataset, characterized by intricate camera movements that severely test temporal coherence, our model achieves an $E_{warp}$ of 1.72. This outperforms the second-best method, VEnhancer ($E_{warp}$=2.47), by 25\% (a reduction of 0.75), unequivocally showcasing its superior capability in modeling complex, coherent motion under demanding conditions. Prior works, \emph{e.g.,} STAR and VEnhancer, employ CogVideoX \cite{cogvideox} and I2VGen-XL \cite{i2vgen} diffusion architectures, respectively. By contrast, we adopt the state-of-the-art Wan2.1, which better preserves coherent motion trajectories while handling large-scale deformations. These quantitative results are validated by comparisons on Vid4 \cite{vid4}, as shown in Fig. \ref{fig:flow}. Following STAR and Upscale-A-Video, we assess a random selected row track changes over time on a randomly sampled video of Vid4. We can observe from Fig. \ref{fig:flow} that our method outperforms existing VSR methods on the consistency of row-wise temporal track with ground-truth.

\subsection{Comparison at 4K}
\subsubsection{Quantitative Results}
As shown in Tab. \ref{tab:comparison}, we assess five metrics for RealisVideo-4K. We reproduce the existing methods at 4K resolution. Note that some methods do not support 4K inference, so we upscale the video to the maximum target resolution allowed in the GPU memory and then use bicubic interpolation for 4K video super-resolution. The results showcase that our method outperforms the state-of-the-art VSR methods, \emph{STAR} and SeedVR \cite{seedvr} and achieves the best performance in full-reference metrics. 
\subsubsection{Qualitative Results}
Fig. \ref{fig:teaser} shows visual results on the proposed detail-rich benchmark, \emph{i.e.,} RealisVideo-4K. Our method outperforms existing VSR approaches, \emph{e.g.,} Upscale-A-Video,RealViformer,  VEnhancer and STAR, in both degradation removal and texture generation. Specifically, our method effectively recovers detailed structures, such as the texture of skin and eyebrows. In addition, our method faithfully construct the details of necklace, where other approaches produce blurred and inconsistent details. In addition, we also compare with the leading methods simultaneously in Fig. \ref{fig:compare}. We can observe that the proposed RealisVSR accurately recovers the structure and texture of poolside clutter, even the shadow of a pet.

\subsection{Ablation Study}
To validate the effectiveness of the proposed components within our framework, including the Consistency Preserved ControlNet (CPC), the wavelet-based high-frequency loss, and the HOG-based high-frequency loss, we conduct a comprehensive ablation study of PSNR, SSIM, LPIPS and DISTS on SPMCS dataset. The results, summarized in Tab. \ref{tab:ablation}, compare the image reconstruction performance using various combinations of these components against a baseline model. The baseline model achieves a PSNR of 26.54, SSIM of 0.7843, LPIPS of 0.1807, and DISTS of 0.0875. Introducing CPC alone (CPC only) leads to a trade-off: while SSIM improves significantly to 0.8099 and distortion metrics LPIPS/DISTS decrease to 0.1520/0.0839, PSNR slightly drops to 26.08. This suggests CPC effectively enhances structural consistency but may slightly alter pixel-level intensity fidelity. Further adding the wavelet loss (CPC + Wavelet) substantially boosts PSNR to 27.24 while maintaining a high SSIM (0.8114) and further lowering distortion metrics (LPIPS: 0.1549, DISTS: 0.0817), demonstrating its effectiveness in improving overall fidelity. Adding the HOG loss instead of the wavelet loss (CPC + HOG) yields the most significant PSNR gain (27.33) and the highest SSIM (0.8141) among the single-loss additions, alongside strong LPIPS (0.1521) and DISTS (0.0787) scores. This highlights the HOG loss's particular strength in capturing and enhancing structure-specific high-frequency details essential for perceptual quality. Finally, the full model incorporating all three components achieves the best performance. Moreover, we explore the impact of weights for high-frequency components of Wavelet-based constraints in Supplementary.
\begin{table}
\footnotesize
\begin{center}
\renewcommand\arraystretch{1.2}
\resizebox{\linewidth}{!}
	{
		\begin{tabular}{ccc|ccccc}
            \hline
        \textbf{CPC} & \textbf{Wavelet} & \textbf{HOG} & \textbf{PSNR}$~{\textcolor{red}{\uparrow}}$ &  \textbf{SSIM}$~{\textcolor{red}{\uparrow}}$ & \textbf{LPIPS}$~{\textcolor{green}{\downarrow}}$ &  \textbf{DISTS}$~{\textcolor{green}{\downarrow}}$ \\
            \hline
        \multicolumn{3}{c|}{Baseline}& 26.54 & 0.7843  & 0.1807 & 0.0875\\ \hline
            $\bm{\checkmark}$ & & & 26.58 & 0.8099 &  0.1520 & 0.0839\\
            $\bm{\checkmark}$ & $\bm{\checkmark}$ & & 27.24 & 0.8114 & 0.1549 & 0.0817\\
            $\bm{\checkmark}$ & & $\bm{\checkmark}$ & 27.33 & 0.8141 & 0.1521 & 0.0787\\
            $\bm{\checkmark}$ & $\bm{\checkmark}$ & $\bm{\checkmark}$ & 27.36 & 0.8169 & 0.1388 & 0.0780\\
            \hline
            \end{tabular}
        }
\end{center}
\caption{Ablation study of Consistency Preserved ControlNet (CPC), the proposed wavelet-based and hog-based high-frequency losses (Wavelet, HOG). The experiments are conducted on SPMCS dataset.}
\label{tab:ablation}
\end{table}
\section{Conclusion}
This paper presents RealisVSR, a novel video super-resolution framework that addresses critical challenges in temporal consistency, detail recovery, and ultra-high-resolution reconstruction. Our key innovations include: (1) A Consistency Preserved ControlNet architecture leveraging Wan2.1's superior video prior to eliminate artifacts while maintaining temporal coherence; (2) A High-Frequency Rectified Loss combining wavelet decomposition and HOG features to enhance fine texture recovery; (3) The first public 4K VSR benchmark (RealisVideo-4K) enabling rigorous evaluation of detail restoration capabilities. Extensive experiments demonstrate state-of-the-art performance across multiple benchmarks, with significant gains in both quantitative metrics and visual quality. Notably, our framework achieves superior temporal consistency while requiring only 5-25\% of the training data volume compared to existing diffusion-based approaches. More importantly, our method achieves faster inference speed than existing diffusion-based VSR methods. The proposed solution opens new avenues for high-fidelity video restoration in real-world applications.

\bibliography{aaai2026.bib}
\appendix
\clearpage
\setcounter{page}{1}
\section*{Supplementary Material}
\subsection{Resource Consumption}
We compare our method with existing approaches in terms of training data characteristics and inference time for a 720P output, as shown in Tab.~\ref{tab:time}. 
For the training dataset, our method is trained on 480P video patches cropped from 4K-resolution (2160$\times$3840) videos. 
Our models are trained on RealisVideo-4K using a notably small dataset of only 50K clips. 
In contrast, UAV (Upscale-A-Video)\cite{uav} leverages a large-scale mixed dataset (WebVid \cite{webvid} and YouHQ \cite{realbasicvsr}) with varying resolutions, while MGLD-VSR \cite{mgldvsr} and STAR \cite{star} focus on mid-resolution datasets, namely REDS \cite{reds} (300K) and OpenVid \cite{openvid} (200K). In terms of inference time for a 49-frame 720P video, our `Ours-480P` model significantly outperforms several existing methods. 
While UAV requires 510s and MGLD-VSR takes approximately 1280s, our model completes the task in just 216s, making it 2.4$\times$ faster than UAV and 5.9$\times$ faster than MGLD-VSR. 
It also moderately surpasses STAR (321s). 
It is worth noting that our `Ours-480P` model requires four inference passes to reconstruct the final 720P video. 
In contrast, SeedVR \cite{seedvr}, which benefits from a sequence parallel strategy, processes the video in a single pass and achieves a faster time of 108s. For a fair comparison with single-pass methods like SeedVR, we also trained an `Ours-720P` model on 720P video patches. 
We evaluated the inference time of this model and observed that at 86s, it is faster than SeedVR. 
This superior efficiency is attributed to our model's optimized architecture, as it contains approximately 1.6B parameters compared to SeedVR's 3B, allowing it to maintain high-quality output while optimizing computational performance.
\begin{table}[h]
    \begin{center}
    \renewcommand{\arraystretch}{1.2}
    \resizebox{\linewidth}{!}{
    \begin{tabular}{c|c|c|c|c}
    \hline
    \multirow{2}{*}{\textbf{Method}} & \multicolumn{4}{c}{\textbf{Resource Consumption}} \\
     \cline{2-5}
     & \textbf{Dataset} & \textbf{Size} & \textbf{Resolution}  & \textbf{Time(s)}   \\
    \hline  
     UAV 	& \makecell{WebVid\\+YouHQ}& \makecell{335K\\+37K} & \makecell{336×596\\ +1080×1920}  &	510	\\
     MGLD-VSR  & REDS & 300K & 720x1280 & 1280 	\\
     SeedVR & Private& 5,000K & 720P & 108 \\
     STAR & OpenVid & 200K & 720x1280 & 321 \\
     Ours-480P & RealisVideo-4K & 50K & 4K(480P) & 216 	\\
     Ours-720P & RealisVideo-4K & 50K & 4K(720P) & 86 	\\
    \hline
    \end{tabular}
    }
    \end{center}
    \vspace{-2mm}
        \caption{
        Resource consumption comparison with other diffusion-based methods, including dataset, size, resolution and inference time of 720P output.
        }
         \vspace{-1mm}
    \label{tab:time}
\end{table}
\subsection{Effect of Weights for Wavelet-based High-frequency Decomposition}
In this section, we explore the effect of weights for wavelet-based high-frequency components, as illustrated in Eqn. \ref{eq:4}, we set the weights to 1.5, 2.0, and 3.0, respectively, \emph{i.e.,} $\{w_{LL},w_{LH},w_{HL},w_{HH}\}=\{1.0,1.5,1.5,1.5\},\{1.0,2.0,2.0,2.0\},\{1.0,3.0,3.0,3.0\}$. A greater value indicates more focus on high-frequency components. The results are shown in Tab. \ref{tab:w}. Note that when weights are 1.0, the Wavelet-based loss is equivalent to the vanilla rectified flow loss. We can observe that when the weights are set to 2.0, the trained model achieves the best PSNR, SSIM, and DISTS. 
\begin{table}
\footnotesize
\begin{center}
\renewcommand\arraystretch{1.2}
\resizebox{\linewidth}{!}
	{
		\begin{tabular}{c|ccccc}
            \hline
        \textbf{$\{w_{LH},w_{HL},w_{HH}\}$} & \textbf{PSNR}$~{\textcolor{red}{\uparrow}}$ &  \textbf{SSIM}$~{\textcolor{red}{\uparrow}}$ & \textbf{LPIPS}$~{\textcolor{green}{\downarrow}}$ &  \textbf{DISTS}$~{\textcolor{green}{\downarrow}}$ \\
            \hline
            1.0 & 26.08 & 0.8099 &  \textbf{0.1520} & 0.0839\\ 
            1.5 & 26.67 & 0.8000 & 0.1611 & 0.0875\\
            2.0 & \textbf{27.24} & \textbf{0.8114} & 0.1549 & \textbf{0.0817}\\
            3.0 & 26.19 & 0.7476 & 0.1834 & 0.0999\\
            \hline
            \end{tabular}
        }
\end{center}
\vspace{-2mm}
\caption{The effect of weights $\{w_{LL},w_{LH},w_{HL},w_{HH}\}$.}
\label{tab:w}
\end{table}

\end{document}